\documentclass[dvips]{acta}
\usepackage{lscape,epsfig}
\usepackage{amssymb}
\usepackage{amsmath}
\usepackage{multirow}

\usepackage{lscape,longtable}
\usepackage{subfigure}
\newcommand{\ApJL}{ApJL}

\begin{document}

\begin{Titlepage}
\Title{Large Magellanic Cloud Cepheids
\\in the ASAS data}

\Author{P.~K~a~r~c~z~m~a~r~e~k$^1$, W.~A.~D~z~i~e~m~b~o~w~s~k~i$^{2,3}$, 
P.~L~e~n~z$^{3}$, P.~P~i~e~t~r~u~k~o~w~i~c~z$^{2}$ and G.~P~o~j~m~a~\'n~s~k~i$^{2}$}
{$^1$ Nicolaus Copernicus University, ul.~Gagarina~11, 87-100~Toru\'n, Poland\\
e-mail: karczmarek@astri.uni.torun.pl\\
$^2$ Warsaw University Observatory, Al.~Ujazdowskie~4, 00-478~Warsaw, Poland\\
$^3$ Copernicus Astronomical Center, ul.~Bartycka~18, 00-716~Warsaw, Poland}

\end{Titlepage}

\Abstract{A catalog of Cepheids in the Large Magellanic Cloud (LMC)
from the ASAS project is presented. It contains data on 65
fundamental mode pulsators with periods longer than about 8 days. The
period-luminosity (PL) relation in the $V$-band does not significantly differ 
from the relation determined by Soszy\'nski \emph{et al.} (2008)
from the OGLE data extended toward longer periods but with much larger
spread. For objects with periods longer than 40 d there is an
evidence for a shallower PL relation. The rates of long-term period variations 
significant at $3\sigma$ level are found only for 7 objects. The rates for 25 objects 
determined with the $1\sigma$ significance are confronted with the values derived from
stellar evolution models. The models from
various sources yield discrepant predictions. Over the whole data
range, a good agreement with measurements is found for certain
models but not from the same source.} {Stars: evolution - Stars:
Cepheids - Large Magellanic Cloud}

\section{Introduction}
Crossing of the Cepheid Instability Strip is a short but important
phase of massive stars evolution. Objects in this phase reveal
periods of radial pulsation, which are valuable observables. The
period-luminosity (PL) relation for Cepheids for nearly a century
plays a crucial role in the cosmic distance ladder. A newer
application of Cepheids is probing massive star evolution through
measurements of the rates of period changes (see, e.g., Pietrukowicz
2001, Turner \emph{et al.} 2006).

Bird \emph{et al.} (2009) emphasize exceptional importance of long
period objects, which they name Ultra Long Period (ULP) Cepheids, in
both applications. Being bright, these objects are seen in distant
galaxies, however the shape of the upper part of the PL relation is not
yet well-established.

Open questions in the theory of massive stars concern, in particular, the rate
of mass loss and the extent of mixing beyond the edge of convective
core. There are related uncertainties in calculations of evolutionary tracks in the H-R diagram and life times in various phases in particular within the Cepheid instability strip.
Here we find considerable differences between results presented by various authors (e.g., Schaerer \emph{et al.} 1993, Fagotto \emph{et al.} 1994, Alibert \emph{et al.} 1999, Bono \emph{et al.} 2000, Pietrinferni \emph{et al.} 2006).

Long period Cepheids in both Magellanic Clouds are being monitored
for nearly 10 years within the ASAS project (Pojma\'nski 2002).
A part of the ASAS data has been already employed in searches for period
changes in a combination with Harvard data by Pietrukowicz (2001).
The long-time data give us a chance to determine the rates of period
change on the basis of the ASAS data alone by means of direct fitting of
measurements. This will be done in the present paper, which is devoted
exclusively to the object located in the LMC.

 An extensive study of
Cepheid period changes in this galaxy, based on the OGLE and MACHO data,
was presented by Poleski (2008) but it was limited to objects with
period below 40 days. The range of periods in the ASAS sample, which
in contrast to that of the OGLE is limited from the bottom, is from 8 to
133 days. There is an overlap enabling a comparison of the results
but also an extension to the interesting ULP range.

In the next section, we present an updated catalog of the LMC
Cepheid in the ASAS data. Section 3 is devoted to the PL relation
based on these data. In Section 4, we present our determination of
period changes and confront them with values calculated for stellar
models.

\begin{figure}[!ht]
\centering
\includegraphics[scale=1.3,clip]{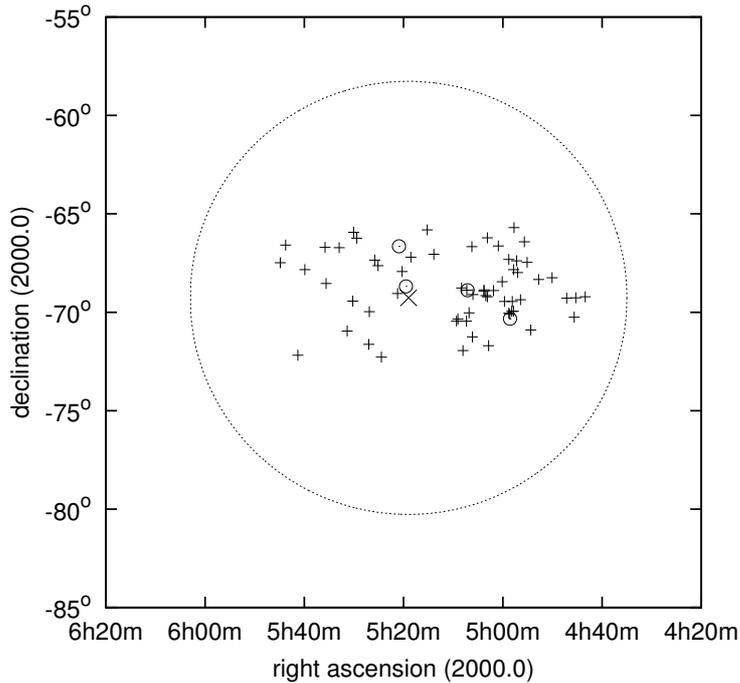}
\caption{LMC Cepheids located inside the circle-shaped area with
$r=11^{\circ}$ from the optical center of the LMC at RA\,=\,$5^{h}
19^{m}$, Dec\,=\,-$69^{\circ} 27'$ (J2000.0 coordinates) (45$^{\circ}$ angled
cross in the middle). Pluses indicate objects classified  in ASAS as
DCEP-FU, open circles show objects originally classified as Miscellaneous or RRab
stars.}
\end{figure}

\section{Inventory of the LMC Cepheids in the ASAS data}

In the ASAS catalog of fundamental mode Cepheids (available in the
internet\footnote{http://www.astrouw.edu.pl/asas/?page=acvs}), we
selected objects located within a radius of 11$^{\circ}$ from the
optical center of the Large Magellanic Cloud (Fig. 1), which is
located at RA\,=\,$5^{h} 19^{m}$, Dec\,=\,-$69^{\circ} 27'$ in J2000.0 coordinates (Subramaniam 2003).

In the specified area, we found 73 stars classified as Cepheids. After a
careful inspection of the light curves, we rejected 5 objects.
Another 7 objects were excluded for their distant positions in the
PL plane from the classical PL relation for Cepheids. Then we looked
for objects located in the same area in the sky with whose original
ASAS classification was either Miscellaneous or RRab stars but
classification as a fundamental mode Cepheid was given as an
alternative. We added 4 objects to our list of Cepheids as they
fulfilled the criteria of light curve shape and location in the PL
plane.

The 65 objects which passed our selection criteria cover the period
range from 8 to 133\,days. Stars with shorter periods are too faint
to be observed by ASAS. Only $V$-band data were employed in this
paper. We have attempted to use the $I$-band data, however the
number of  measurements in this band was insufficient to determine
meaningful light curves. Figs. 2 and 3 present the $V$-band light
curves of all objects except of ASAS 050601-6906.3, which is very similar to ASAS 050346-6852.6. 
The light curves are repeated twice for clarity. We see no trend in the 
behavior of amplitudes with period.
The lack of an apparent trend may be in part blamed to blending.
Although the amplitudes of the stars are uncertain, it does not
affect the accuracy of period changes which are of our main interest
in this paper.

Selected data on all confirmed Cepheids are listed in Table 1. Data
in consecutive columns of the table give: ASAS star identification,
 OGLE and/or Harvard identifications (if available), mean $V$
magnitude, amplitude, period, dimensionless rate of period change,
signal-to-noise ratio (S/N) for the rate, and number of data points.

\newpage
\begin{figure}[!ht]
\centering
\includegraphics[width=\textwidth,clip] {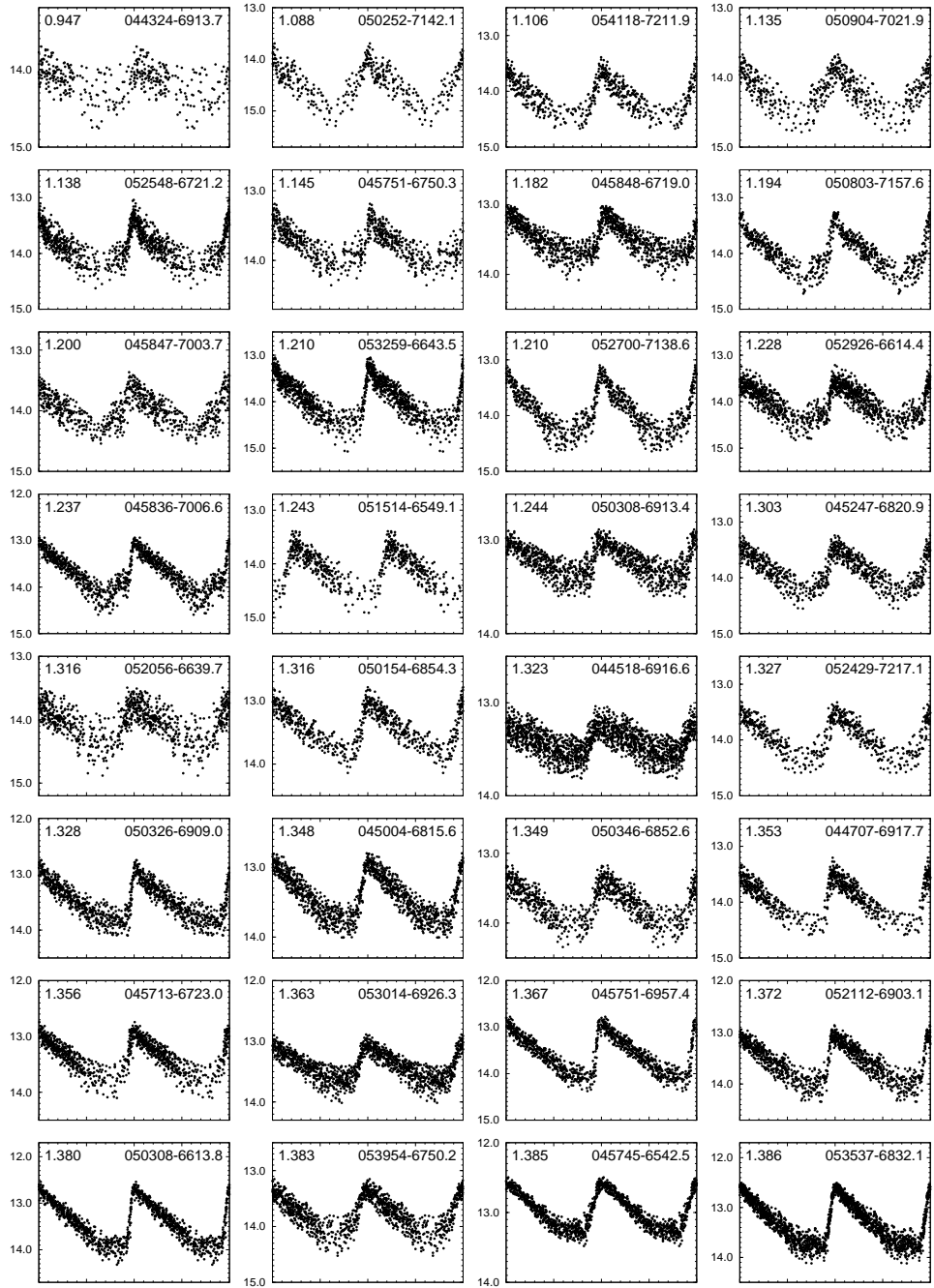}
\caption{$V$-band light curves of the ASAS Cepheids with periods in the
range of 8-24 days. In each graph top left number shows the
logarithm of period in days and top right the ASAS ID of the star.
The light curves are repeated twice for clarity.}
\end{figure}

\begin{figure}[!ht]
\centering
\newpage
\includegraphics[width=\textwidth,clip] {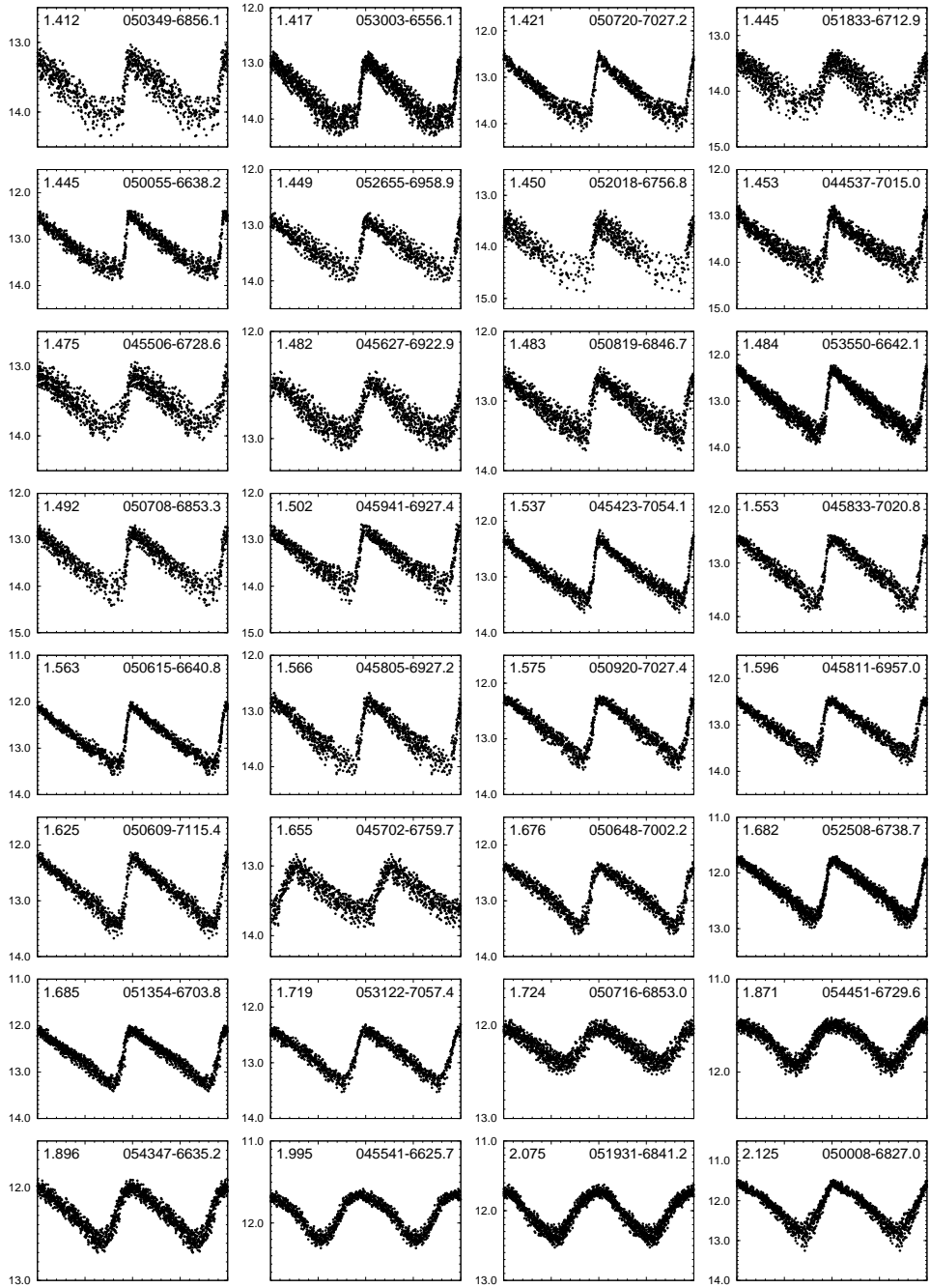}
\caption{The same as Fig. 2 but period range is 25-133 days.}
\end{figure}

\vspace*{5mm}
\begin{footnotesize}
\begin{longtable}{p{2cm} p{4.4cm} p{17mm} p{5mm} p{5mm} r p{4mm} r}\\
\caption{List of 65 ASAS Cepheids in the Large Magellanic Cloud} \\
\noalign{\smallskip} \hline \noalign{\smallskip}
\multirow{2}{*}{ASAS ID} & \multirow{2}{*}{OGLE/Harvard ID} & \hspace*{2mm} $\langle V \rangle \pm \Delta V$ & Amp. & $\log(P)$  & $\dot{P}$ \hspace*{3mm} & \multirow{2}{*}{S/N} & \multirow{2}{*}{$N$}   \\
 & & \hspace*{4mm} [mag] & [mag] & \hspace*{1mm} [d] & $\times 10^{-6}$ & & \\
\noalign{\smallskip} \hline \noalign{\smallskip}
050008-6827.0   &   HV883                    &  12.18  $\pm$   0.11   & 1.41 & 2.125 &  -48.1  & 0.62    &   764 \\
051931-6841.2   &   HV2447                   &  12.04  $\pm$   0.06   & 0.67 & 2.075 &   24.0  & 0.46    &  1010 \\
045541-6625.7   &   HV5497                   &  11.90  $\pm$   0.05   & 0.66 & 1.995 &  232.7  & 6.48    &   661 \\
054347-6635.2   &   HV2827                   &  12.28  $\pm$   0.06   & 0.63 & 1.896 &  161.4  & 6.94    &   833 \\
054451-6729.6   &   -                        &  11.68  $\pm$   0.05   & 0.47 & 1.871 &  -22.7  & 1.02    &   953 \\
050716-6853.0   &   OGLE-LMC-CEP-0992        &  12.22  $\pm$   0.06   & 0.47 & 1.724 &   -4.5  & 0.22    &   776 \\
053122-7057.4   &   OGLE-LMC-CEP-2253/HV2622 &  12.88  $\pm$   0.07   & 0.86 & 1.719 &    6.0  & 0.54    &   580 \\
051354-6703.8   &   OGLE-LMC-CEP-1290/HV2369 &  12.68  $\pm$   0.08   & 1.14 & 1.685 &   56.1  & 7.24    &   784 \\
052508-6738.7   &   HV953                    &  12.29  $\pm$   0.06   & 1.08 & 1.682 &   11.3  & 2.09    &  1163 \\
050648-7002.2   &   OGLE-LMC-CEP-0966/HV900  &  12.89  $\pm$   0.08   & 1.09 & 1.676 &   80.9  & 6.90    &   606 \\
045702-6759.7   &   OGLE-LMC-CEP-0461/HV877  &  13.38  $\pm$   0.10   & 0.92 & 1.655 &  -15.3  & 0.95    &   582 \\
050609-7115.4   &   OGLE-LMC-CEP-0943/HV2338 &  12.84  $\pm$   0.08   & 1.22 & 1.625 &   32.1  & 4.64    &   615 \\
045811-6957.0   &   OGLE-LMC-CEP-0512/HV2257 &  13.07  $\pm$   0.09   & 1.35 & 1.596 &   -2.3  & 0.30    &   615 \\
050920-7027.4   &   OGLE-LMC-CEP-1113/HV909  &  12.83  $\pm$   0.08   & 1.09 & 1.575 &  -20.9  & 3.64    &   661 \\
045805-6927.2   &   OGLE-LMC-CEP-0510/HV879  &  13.43  $\pm$   0.10   & 1.28 & 1.566 &    3.1  & 0.39    &   551 \\
050615-6640.8   &   OGLE-LMC-CEP-0945/HV2294 &  12.78  $\pm$   0.07   & 1.27 & 1.563 &    1.1  & 0.26    &   657 \\
045833-7020.8   &   OGLE-LMC-CEP-0528/HV881  &  13.14  $\pm$   0.09   & 1.27 & 1.553 &    3.2  & 0.54    &   628 \\
045423-7054.1   &   OGLE-LMC-CEP-0328/HV873  &  12.92  $\pm$   0.07   & 1.23 & 1.537 &   -4.5  & 0.92    &   690 \\
045941-6927.4   &   OGLE-LMC-CEP-0590/HV882  &  13.45  $\pm$   0.11   & 1.32 & 1.502 &    1.5  & 0.23    &   588 \\
050708-6853.3   &   OGLE-LMC-CEP-0986/HV899  &  13.54  $\pm$   0.13   & 1.32 & 1.492 &   11.6  & 1.64    &   515 \\
053550-6642.1   &   HV1002                   &  13.06  $\pm$   0.10   & 1.34 & 1.484 &   13.0  & 5.45    &  1196 \\
050819-6846.7   &   OGLE-LMC-CEP-1058/HV904  &  13.08  $\pm$   0.09   & 0.88 & 1.483 &   -2.9  & 0.45    &   780 \\
045627-6922.9   &   OGLE-LMC-CEP-0434/HV875  &  12.75  $\pm$   0.07   & 0.66 & 1.482 &  -13.0  & 1.56    &   661 \\
045506-6728.6   &   OGLE-LMC-CEP-0367/HV872  &  13.52  $\pm$   0.12   & 0.89 & 1.475 &    5.3  & 0.61    &   620 \\
044537-7015.0   &   OGLE-LMC-CEP-0068/HV8036 &  13.62  $\pm$   0.12   & 1.33 & 1.453 &   -2.7  & 0.61    &   823 \\
052018-6756.8   &   OGLE-LMC-CEP-1632/HV934  &  14.09  $\pm$   0.16   & 1.05 & 1.450 &    2.5  & 0.23    &   421 \\
052655-6958.9   &   OGLE-LMC-CEP-2019/HV2540 &  13.41  $\pm$   0.11   & 0.95 & 1.449 &   -3.0  & 0.37    &   515 \\
050055-6638.2   &   OGLE-LMC-CEP-0654/HV225  &  13.17  $\pm$   0.11   & 1.31 & 1.445 &   -4.8  & 1.07    &   600 \\
051833-6712.9   &   HV929                    &  13.87  $\pm$   0.14   & 0.86 & 1.445 &    5.9  & 0.84    &   687 \\
050720-7027.2   &   OGLE-LMC-CEP-0999/HV902  &  13.34  $\pm$   0.10   & 1.33 & 1.421 &    1.7  & 0.48    &   603 \\
053003-6556.1   &   HV12815                  &  13.59  $\pm$   0.12   & 1.03 & 1.417 &    3.1  & 0.86    & 1013  \\
050349-6856.1   &   OGLE-LMC-CEP-0821/HV889  &  13.72  $\pm$   0.13   & 1.02 & 1.412 &   -9.4  & 1.71    &   515 \\
053537-6832.1   &   OGLE-LMC-CEP-2504/HV1003 &  13.33  $\pm$   0.11   & 1.20 & 1.386 &   -1.3   & 0.48    &   947 \\
045745-6542.5   &   HV6098                   &  12.99  $\pm$   0.07   & 0.88 & 1.385 &    2.1   & 0.75    &   625 \\
053954-6750.2   &   OGLE-LMC-CEP-2832/HV1013 &  13.83  $\pm$   0.14   & 0.95 & 1.383 &   -0.2   & 0.03    &   576 \\
050308-6613.8   &   HV886                    &  13.47  $\pm$   0.12   & 1.41 & 1.380 &    4.0   & 1.25    &   582 \\
052112-6903.1   &   OGLE-LMC-CEP-1677/HV938  &  13.64  $\pm$   0.13   & 1.03 & 1.372 &    1.7   & 0.33    &   572 \\
045751-6957.4   &   OGLE-LMC-CEP-0501/HV878  &  13.63  $\pm$   0.12   & 1.42 & 1.367 &    2.5   & 0.71    &   512 \\
053014-6926.3   &   OGLE-LMC-CEP-2176/HV984  &  13.43  $\pm$   0.12   & 0.65 & 1.363 &   11.0   & 2.14    &   658 \\
045713-6723.0   &   OGLE-LMC-CEP-0467/HV876  &  13.44  $\pm$   0.13   & 1.16 & 1.356 &    1.5   & 0.33    &   506 \\
044707-6917.7   &   OGLE-LMC-CEP-0079/HV1    &  14.05  $\pm$   0.13   & 1.03 & 1.353 &   15.2   & 2.84    &   387 \\
050346-6852.6   &   OGLE-LMC-CEP-0819/HV2291 &  13.76  $\pm$   0.13   & 0.81 & 1.349 &    0.5   & 0.08    &   448 \\
045004-6815.6   &   OGLE-LMC-CEP-0147/HV12446&  13.44  $\pm$   0.11   & 0.98 & 1.348 &   -1.3   & 0.35    &   627 \\
050326-6909.0   &   OGLE-LMC-CEP-0801/HV2292 &  13.52  $\pm$   0.13   & 1.01 & 1.328 &   -4.6   & 1.28    &   560 \\
052429-7217.1   &   HV12804                  &  14.02  $\pm$   0.14   & 0.95 & 1.327 &  -10.7   & 2.00    &   297 \\
050601-6906.3   &   -						 &  14.10  $\pm$   0.16   & 1.34 & 1.325 &    2.2   & 0.45    &   369 \\
\hline \noalign{\smallskip}
\multirow{2}{*}{ASAS ID} & \multirow{2}{*}{OGLE/Harvard ID} & \hspace*{2mm} $\langle V \rangle \pm \Delta V$ & Amp. & $\log(P)$  & $\dot{P}$ \hspace*{3mm} & \multirow{2}{*}{S/N} & \multirow{2}{*}{$N$}   \\
 & & \hspace*{4mm} [mag] & [mag] & \hspace*{1mm} [d] & $\times 10^{-6}$ & & \\
\noalign{\smallskip} \hline \noalign{\smallskip}
044518-6916.6   &   OGLE-LMC-CEP-0063        &  13.43  $\pm$   0.11   & 0.57 & 1.323 &    3.5   & 0.60    &   911 \\
052056-6639.7   &   -						 &  14.10  $\pm$   0.17   & 0.93 & 1.316 &   -3.7   & 0.43    &   443 \\
050154-6854.3   &   OGLE-LMC-CEP-0712/HV885  &  13.49  $\pm$   0.11   & 1.02 & 1.316 &   -2.0   & 0.46    &   357 \\
045247-6820.9   &   OGLE-LMC-CEP-0249/HV11   &  13.92  $\pm$   0.13   & 0.93 & 1.303 &    2.7   & 0.74    &   419 \\
050308-6913.4   &   OGLE-LMC-CEP-0787/HV2288 &  13.24  $\pm$   0.09   & 0.50 & 1.244 &    0.0   & 0.01    &   580 \\
051514-6549.1   &   HV2888                   &  14.21  $\pm$   0.15   & 1.13 & 1.243 &   -4.3   & 1.05    &   317 \\
045836-7006.6   &   OGLE-LMC-CEP-0535/HV2261 &  13.75  $\pm$   0.15   & 1.40 & 1.237 &   -0.1   & 0.03    &   523 \\
052926-6614.4   &   HV2580                   &  14.10  $\pm$   0.17   & 0.98 & 1.228 &    0.5   & 0.17    &   634 \\
052700-7138.6   &   OGLE-LMC-CEP-2023/HV2549 &  13.96  $\pm$   0.14   & 1.10 & 1.210 &    3.2   & 1.34    &   420 \\
053259-6643.5   &   HV2667                   &  14.07  $\pm$   0.16   & 1.11 & 1.210 &    5.8   & 2.61    &   651 \\
045847-7003.7   &   OGLE-LMC-CEP-0545/HV2262 &  14.03  $\pm$   0.16   & 0.93 & 1.200 &    3.0   & 0.69    &   379 \\
050803-7157.6   &   HV1276                   &  14.02  $\pm$   0.15   & 1.09 & 1.194 &    1.9   & 0.70    &   379 \\
045848-6719.0   &   OGLE-LMC-CEP-0546/HV2249 &  13.52  $\pm$   0.12   & 0.70 & 1.182 &   -4.7   & 1.87    &   604 \\
045751-6750.3   &   OGLE-LMC-CEP-0500/HV2244 &  13.82  $\pm$   0.14   & 0.94 & 1.145 &   -1.4   & 0.43    &   360 \\
052548-6721.2   &   HV955                    &  13.95  $\pm$   0.15   & 0.97 & 1.138 &   -0.7   & 0.31    &   543 \\
050904-7021.9   &   OGLE-LMC-CEP-1100/HV2352 &  14.26  $\pm$   0.14   & 0.88 & 1.135 &    3.5   & 1.08    &   262 \\
054118-7211.9   &   OGLE-LMC-CEP-2922/HV12839&  14.16  $\pm$   0.14   & 0.91 & 1.106 &    2.6   & 1.04    &   297 \\
050252-7142.1   &   HV12745                  &  14.58  $\pm$   0.18   & 1.21 & 1.088 &   -4.6   & 1.70    &   201 \\
044324-6913.7   &   OGLE-LMC-CEP-0046/HV12717&  14.18  $\pm$   0.17   & 0.59 & 0.947 &    3.7   & 0.93    &   198 \\
\hline
\end{longtable}
\end{footnotesize}

\section{Period-luminosity relation}

\begin{figure}[!ht]
\centering
\includegraphics[scale=1.3,clip]{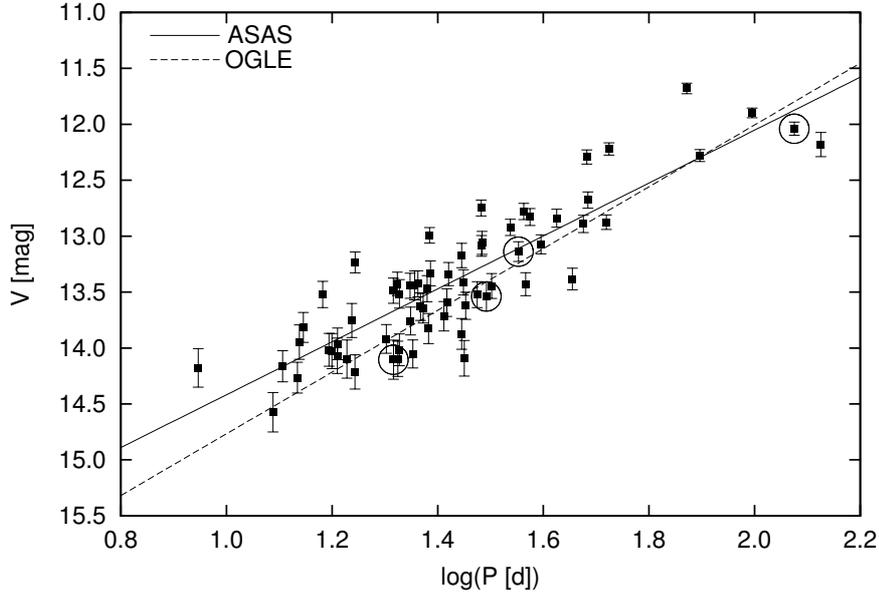}
\caption{Period-luminosity relation for the ASAS Cepheids. The four objects added to ASAS are shown with the encircled symbols. The solid line corresponds to the least square fit (see Eq. (1)). The dashed line shows the same for the OGLE Cepheids (see Eq. (2)), as determined by Soszy\'nski \emph{et al.} (2008).}
\end{figure}

With the data given in Table 1, we constructed the PL diagram in
$V$-band, which is shown in Fig. 4. Errors in period measurements
are very small (within the symbols). The linear least squares fit
yields
\begin{equation}
V_{\rm ASAS}  = -2.366(\pm 0.166) \log P + 16.784(\pm 0.239), \quad \sigma = 0.31 \rm ~ mag.
\end{equation}
This may be compared with the PL relation for single-mode Cepheids determined by Soszy\'nski \emph{et al.} (2008)
on the basis of the OGLE data
\begin{equation}
V_{\rm OGLE}  = -2.762(\pm 0.022) \log P + 17.530(\pm 0.015), \quad \sigma = 0.23 \rm ~ mag.
\end{equation}
The lines corresponding to these two relations are shown in Fig. 4.
The four added objects are well within 3$\sigma$ of the ASAS fit which was
the argument for including them to the sample.

We may see that our relation is not in a significant conflict with the
extension of the OGLE PL relation, which was based on the sample
limited to $\log(P) < 1.7$. The slopes differ by less than $3\sigma$.
The scatter of the ASAS luminosities is much larger, which in part is due to much
smaller size of our sample and in part due to blending. The shallower
slope of the ASAS relation may result from statistical bias connected
with different magnitude ranges in the two samples but it is well
possible that the difference reflects a real nonlinearity of the PL relation. 
In fact, a significant flattening of this relation in the long period
range has been suggested by Bird \emph{et al.} (2009).
For their sample of the ULP Cepheids, they find the slope $-1.09
\pm 0.94$ with $\sigma$ = 0.40 mag. The lower period limit for these 
objects is $\log (P)=1.9$. Our sample contains only three objects above
this limit. In the range extended down to $\log(P) > 1.6$ we have
12 objects. For this sample the slope of $-1.95 \pm 0.65$ 
with $\sigma$ =0.38 mag, which is not in contradiction with the
assessment quoted by Bird \emph{et al.}.

\section{Search for evolutionary period changes}

Certainly most of long-period Cepheids are objects burning helium in
their convective core. Measuring rate of period changes for these
objects is important because the evolution is still not fully
understood. The difficulties in modeling stars in this phase of stellar
evolution were first discussed by Paczy\'nski (1970). The problem was subsequently
studied in great details  by Lauternborn \emph{et al.} (1971), whose results
were described in a separate chapter of the Kippenhahn and Weigert
(1990) monograph.  Nonetheless, considerable and difficult to
explain differences between evolutionary tracks calculated by
different authors are present to these days. The crucial problem is
the occurrence and extent of the blue loops in this phase
of evolution.  We still do not know which of stellar and/or modeling parameters play
the role. Most recent models from BaSTI library (Pietrinferni
\emph{et al.} 2006) do not help in answering this question.
 Uncertainties in modeling are best reflected in the rate of stellar
period changes.
\begin{figure}[!ht]
\centering
\includegraphics[scale=1.3,clip]{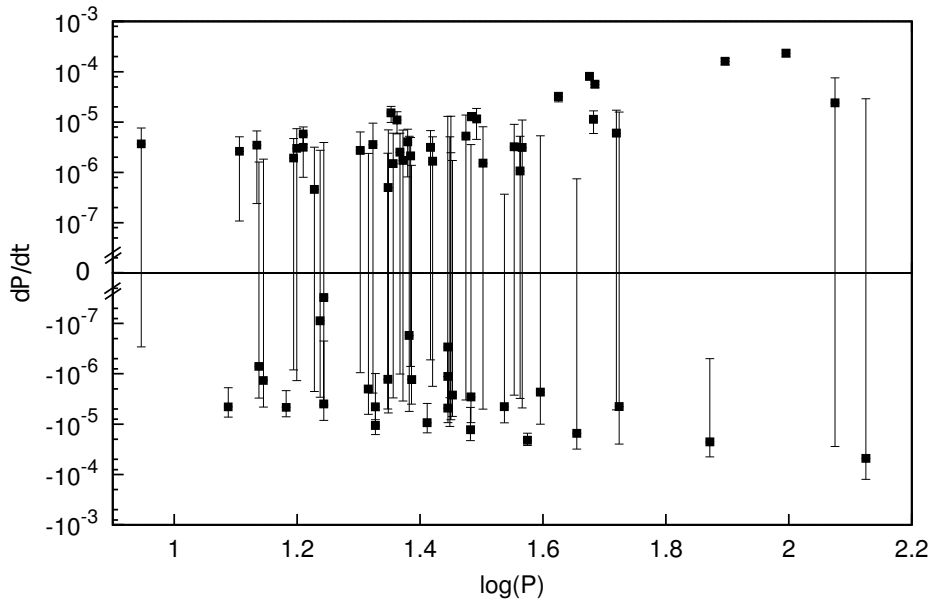}
\caption{Rates of period change for the whole ASAS sample of LMC Cepheids.
The $1\sigma$ error bars represent uncertainty of the fit of measurements to Eq.(3).}
\end{figure}
\subsection{The method and results}
To conduct our investigation on period changes, a modified version of \textsf{Period04} (Lenz and Breger 2005) was used.
Fortunately, the observed period changes are small and do not harm the effectiveness of frequency detection by Fourier techniques.
 The formula used for least squares fitting, however, was modified to include a parabolic frequency change,
\begin{equation}
  f(t) = A_0 + \sum_{i=1}^{n} A_i \sin( 2\pi(\gamma t^2 + \nu_i t + \phi_i)),
  \label{eq:lsfit}
\end{equation}
where $A_0$ is the zero point in amplitude, $n$ is the number of
terms, $\gamma$ is the frequency change parameter and $A_i$,
$\nu_i$, and $\phi_i$ are amplitude, frequency and phase of the term $i$.
Nonlinear least squares fits require good initial parameters;
otherwise, the algorithm may get trapped in a local minimum of the
$\chi^2$ hypersurface. The discrete Fourier transform provides an
initial guess for the frequencies and amplitudes. To obtain good
initial values for the other parameters we used the following
sequential approach: first we adopted the frequency as given by the
Fourier routine, assumed $\gamma=0$ and improved only the amplitudes
and phases by means of a least squares fit. Based on this solution
we determined a better initial value for the frequency change
parameter $\gamma$. The modified \textsf{Period04} version computed
$\chi^2$ for different test values of $\gamma$ ranging within a
predefined interval from $-10^{-6}$ to $+10^{-6}$. The
$\gamma$ value with the lowest $\chi^2$ was then used as initial
value to compute the full least squares solution according to the
formula given in Eq. (3).

To give an estimate for the confidence limits of the parameters, \textsf{Period04} computed the parameter uncertainties
as a byproduct of the least squares fit and, as a second independent option, through Monte Carlo simulations.
 Finally, the frequency change parameter, $\gamma$, was transformed to the period change $dP/dt$ as given $dP/dt = -2 \gamma/\nu^2$, taking into account error propagation.

The fact that the period change can be determined for any Cepheid is
undeniable advantage of this method. Unfortunately,
as we may see in Fig. 5, in many cases uncertainties of the $dP/dt$ are huge.
 In the sample of 65 Cepheids only for 7 objects the measured rates are significant at
the 3$\sigma$ level (S/N=3). For comparison with rates of evolutionary period changes calculated from stellar
models we will use the values for 25 objects with S/N>1.

In Fig. 6 we compare uncertainties of the period change rates for all 65 objects as determined with
\textsf{Period04} code with the values based on Pilecki's \emph{et al.} (2007) approximate expression
for the uncertainty of the period change rate,
\begin{equation}
\sigma_{\rm est} \approx \frac{12 P^2_{\rm sin}}{N^{1/2}T^2} \frac{\sigma}{A},
\end{equation}
which is valid in the case of a star with the sinusoidal light curve observed $N$ times, 
uniformly distributed over time $T$ with a photometric
error $\sigma$, period of a sinusoid $P_{\rm sin}$ and full-amplitude $A$.
We see a good correlation but $\sigma_{\rm P04}$ is 1.2 to 2 times greater than $\sigma_{\rm est}$.

\begin{figure}[!ht]
\centering
\includegraphics[scale=1.3,clip]{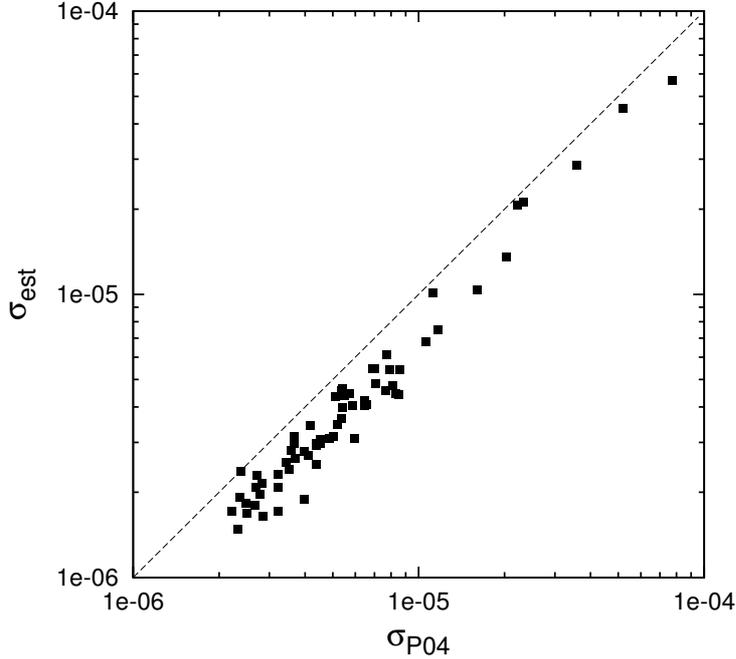}
\caption{Comparison of uncertainties computed with \textsf{Period04} with those calculated with Eq.~(4).}
\end{figure}

We should stress that the uncertainties may include not only the measurement errors but possibly also the
effect of fast period variations arising from nonlinear effects in stellar pulsation.
It is impossible to disentangle these two effects in individual cases.
Poleski (2008), who studied period changes in the LMC Cepheids in the OGLE and MACHO data, found fast variation only in 18 percent of fundamental mode pulsators. Thus, it seems unlikely that such
variations are main source of the large uncertainties.

For some of our objects, we could compare our rates of period change
with the results published by Pietrukowicz (2001) and Poleski (2008).
Pietrukowicz (2001) based his determination of period changes on differences between the ASAS and Harvard
periods for 19 LMC Cepheids. The advantage of his approach is the long-time base but the use of period rather
than phase data is the price. The comparison of the rates determined in this way with our values is shown in Fig. 7.
The agreement is satisfactory. Only in one case the difference exceeds $2\sigma$.

\begin{figure}[!ht]
\centering
\includegraphics[scale=1.17,clip] {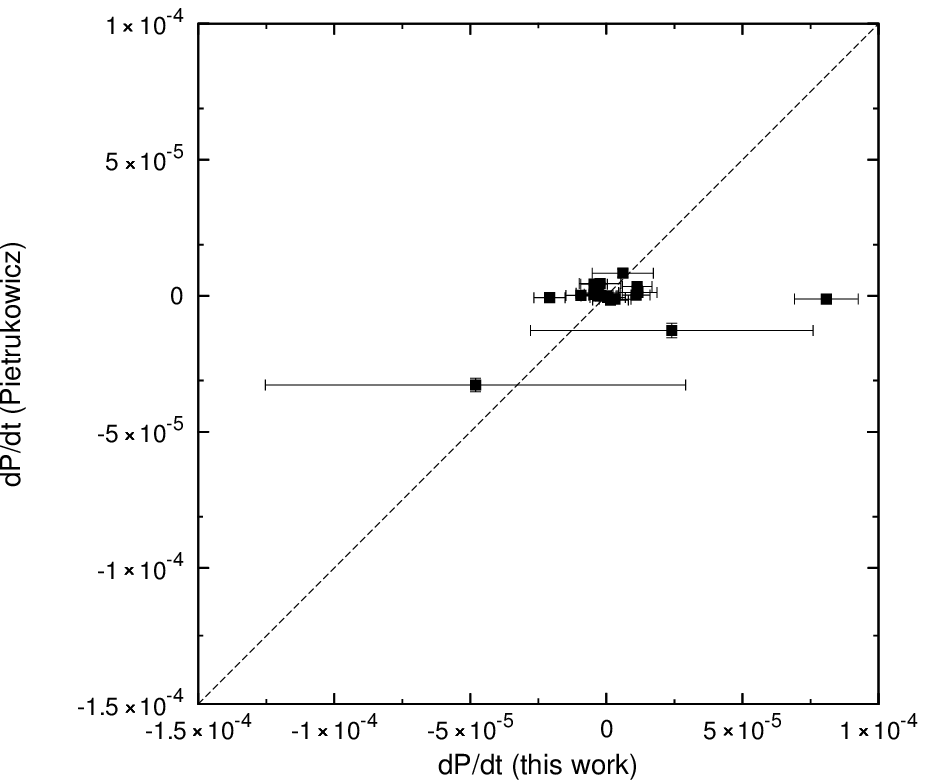}
\caption{Comparison of period rate changes determined in this work with the values obtained by Pietrukowicz (2001).}
\end{figure}

Figure 8 compares rates of period changes determined in the present work with the values obtained by Poleski (2008)
for the OGLE data. In his determinations, he relied on the traditional O-C diagrams method and on
Fourier parameter fitting, similar to ours. From his large sample, we used only rates for 7 overlapping objects with S/N>3.
The agreement is again fairly good.
In the two cases where larger differences between the results are seen, the uncertainty in the OGLE rates
might have been underestimated (Poleski, priv. com.).

As all our predecessors, we assume that the linear period changes which we determined
for the the selected 25 Cepheids yield a realistic assessment of period changes resulting solely from evolutionary
changes on stellar parameters. It should be stressed, however, that
it is an assumption because nonlinear effects in stellar pulsations are not yet understood well enough to
exclude pulsational origin of very slow period changes.
We now proceed to compare the rates we determined with the  values calculated for different sequences of stellar models.
The comparison is done in the $(\log P,{\rm d}P/{\rm d}t)$ plane. Certainly, with the use of additional data, the comparison would be more constraining but ASAS do not provide any with an adequate precision.

\begin{figure}[!ht]
\centering
\includegraphics[scale=1.17,clip] {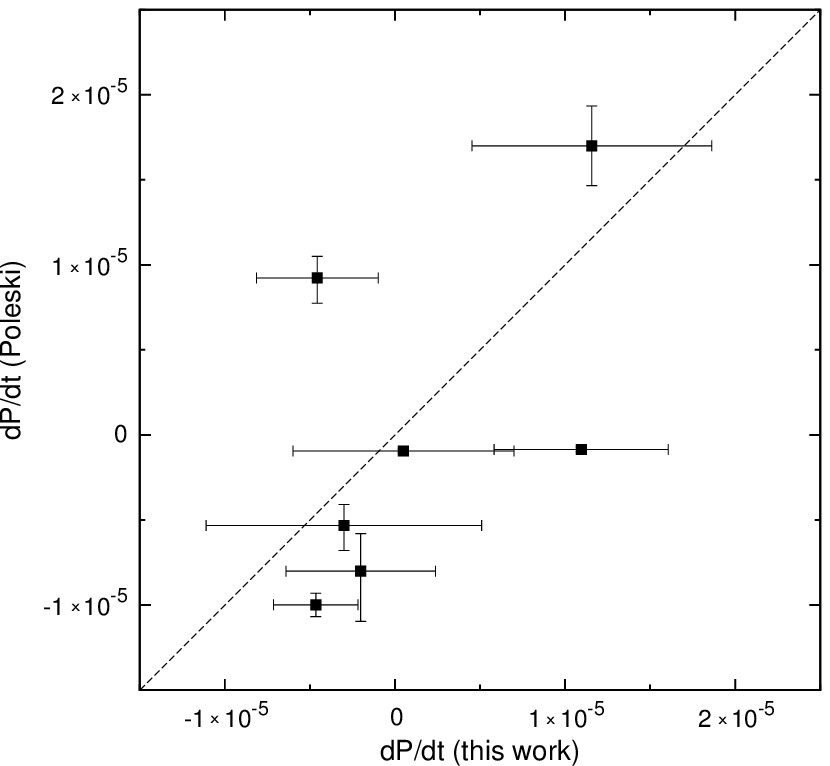}
\caption{Comparison of period rate changes determined in this work with the values obtained by Poleski (2008).}
\end{figure}

\subsection{Comparison with predictions from stellar models}

\begin{figure}[!ht]
\centering
\includegraphics[scale=1.17,clip] {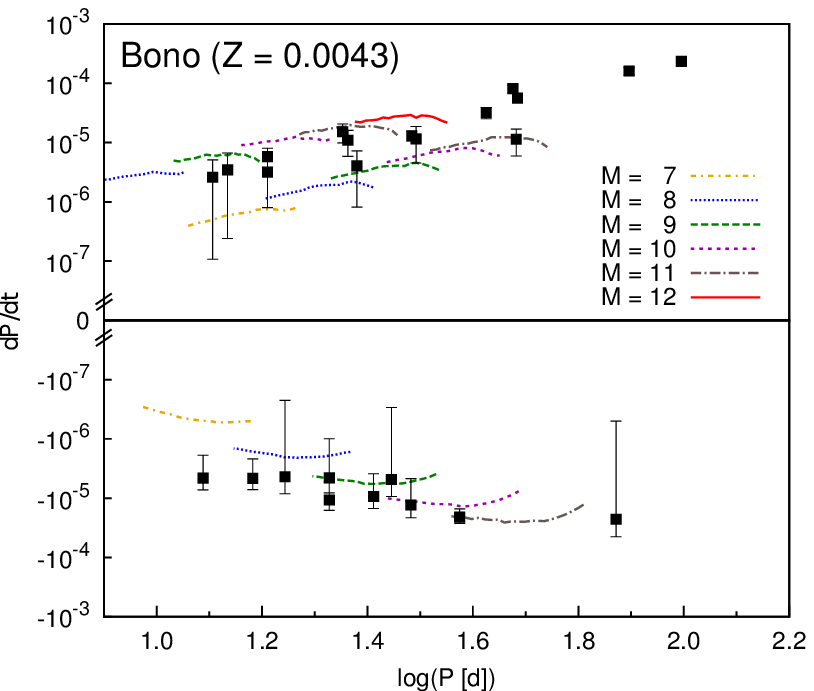} 
\vspace*{0.4cm}
\includegraphics[scale=1.17,clip] {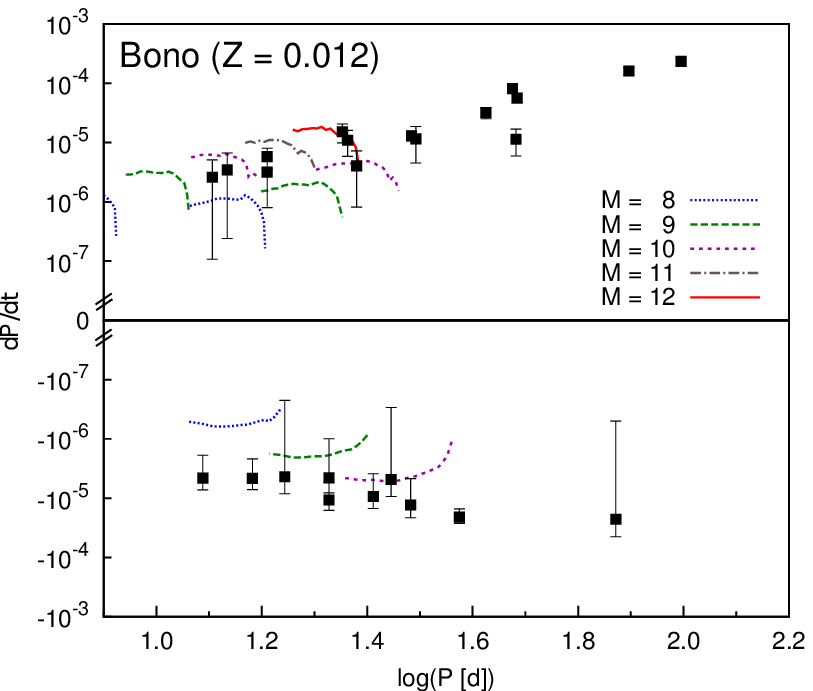}
\caption{Comparison of the measured rates of period change with rates calculated from
Bono \emph{et al.} (2000) models at two indicated metallicity parameters, $Z$, and indicated
initial masses, $M$, in solar units. The third crossing may be distinguished from the first crossing by the lower
rates at specified periods and the longer periods at specified mass. At $M=12 M_{\odot}$ there is only
one crossing.}
\end{figure}

In Fig. 9 rates of period changes determined for
the ASAS sample of LMC Cepheid are compared with the calculated
values obtained from Bono et al. (2000) models at two metal abundance
parameters, $Z$, around the value of $Z=0.008$, which is typically assumed for
the young LMC objects. We should note that the positive measured rates are between those
calculated for the first and third crossing of the instability strip for both values of $Z$.
Since the relative chance for capturing an object during the first crossing is expected much
lower than during the third crossing, this may suggest that the calculated rate for
the third crossing is somewhat low. As for the negative rates, the agreement with the calculated rates
at $Z=0.0043$ is good but at $Z=0.012$ the calculated rates are again somewhat lower than measured.
Pietrukowicz (2001), who also compared his determinations with rates inferred from Bono \emph{et al.}
models, concludes that this comparison revealed "a crude agreement".
Unfortunately, the calculated numbers cover only the short period part of our sample.

To cover the whole range of periods, we used data from
evolutionary tracks calculated by Fagotto \emph{et al.} (1994) and 
Schaerer \emph{et al.} (1993). The data do not include
pulsation periods and, thus, we calculated them with our code for
envelope models with the surface parameters taken from the tracks. The
models are rather scarce in time, therefore we determined only ages and periods
interpolated at the ends of the instability strip crossings. We approximated the
values of effective temperatures at the blue and red ends of the
instability strip with the linear relations, $\log T_{\rm eff,
blue}=3.803-0.045[\log(L/L_{\odot})-3]$ and $\log T_{\rm eff,
red}=3.740-0.060[\log(L/L_{\odot})-3]$,
respectively, which are based on nonlinear pulsation models calculated by  Smolec and Moskalik (2008).
For these models, only one rate for each crossing is given in Fig.~10.
The bars at a given rate extend over the period ranges for the corresponding instability strips.

The models were calculated with the same metal abundance parameter,
$Z=0.008$. The initial He abundance adopted by Fagotto \emph{et al.}
models is by about 5 percent lower than by Schaerer \emph{et al.}
There are differences in the adopted mass loss rate and overshooting
from convective cores. In a consequence the period ranges in the
corresponding instability strips at the same mass are somewhat
shorter in the former models. The rates of period decrease during
the second crossing are similar and, especially in the short period
range, much lower than we determined from the ASAS data. Let us
recall that in this range the measured rates agree quite well with
calculated with Bono \emph{et al.} (2000) models. As we may see in
Fig. 10, the rates of period increase calculated by Fagotto \emph{et
al.} and Schaerer \emph{et al.} differ quite a lot. In the first
case, the mean rate of period change is significantly lower during
the third than the first crossing, except at $M=15 M_{\odot}$, when
the opposite is true. In the second case, the rates are very similar
during these two crossing except at $M=15 M_{\odot}$, when the rate
during the third crossing is significantly lower. Our rates of
period increase are satisfactorily  reproduced with Schaerer
\emph{et al.} (1993) models at $\log P\le 1.7$ but at longer periods
Fagotto \emph{et al.} (1994) models do much better. The measured
rates of period decrease are somewhat faster than predicted by
models from both sources.

\begin{figure}[!ht]
\centering
\includegraphics[scale=1.17,clip] {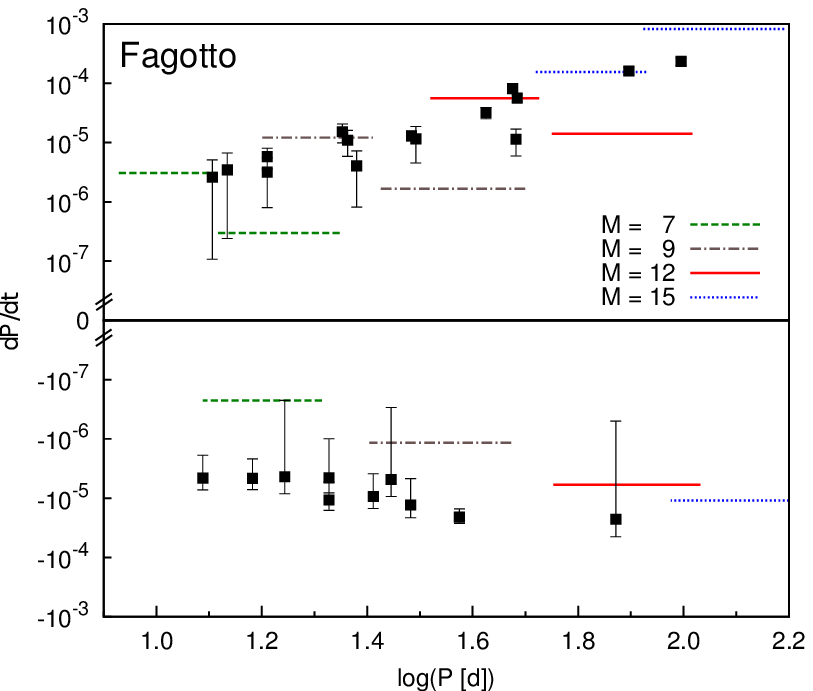}
\includegraphics[scale=1.17,clip] {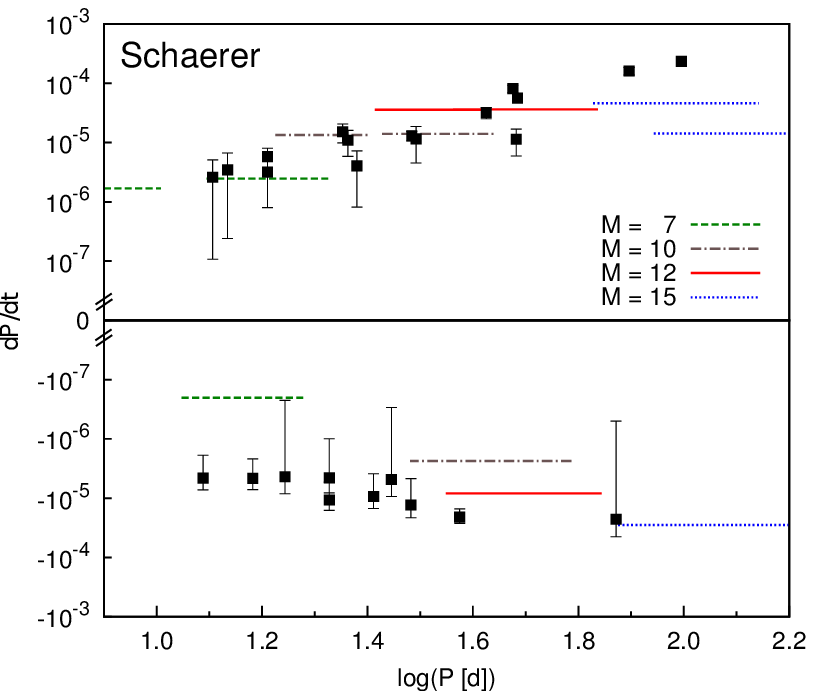}
\caption{Comparison of the measured rates of period change with rates calculated from
Fagotto \emph{et al.} (1994) and Schaerer \emph{et al.} (1993) models at indicated
initial masses, $M$, in solar units.}
\end{figure}

\section{Conclusions}

We updated the list of Large Magellanic Cloud Cepheids in the ASAS
sample and recalculated their pulsation characteristics and mean
magnitudes in the $V$-band. The current list, which is given in Section 2, 
contains 65 confirmed objects covering the period range
from 8 to 133 days. With these data, we determined the linear PL
relation.

Within its large uncertainty, the PL relation from  ASAS data does not significantly differ from 
that determined based on the OGLE data (Soszy\'nski
\emph{et al.} 2008) linearly extended towards longer periods.
 For the sample of 12 objects with period longer than 40\,days we found a flatter
dependance which agrees with that determined by Bird \emph{et al.}
(2009). The large uncertainty is presumably caused mainly by
reddening. Unfortunately the scarce data in the $I$-band did not
allow to determine the reddening-free relation employing the
Wesenheit index. Hopefully it will be possible in the future and
this will help in the interpretation of period changes.

We tried to determine the rates of evolutionary period changes for all
65 objects. However, only for 25 we derived values significant at
1$\sigma$ level. The uncertainty may arise from measurement errors but
it may arise also from non-evolutionary period variations.  We
argued that the first cause is dominant. Both, negative and positive rates were found. 
For some of the objects, we could compare the rates
with earlier determinations and we find a reasonable agreement.

The derived rates were also compared with the values calculated for
stellar models. We noted that the models from different sources yield very different values at
similar parameters. The negative rates in the short period range are well reproduced
by models in the second crossing phase calculated by Bono \emph{et al.} (2000).  The positive
rates in the short-period range are in good agreement with Schaerer \emph{et al.} (1993) models
but in the long-period range the agreement with Fagotto \emph{et al.} (1994) models
is much better.
\Acknow{We thank Radek Poleski for helpful information and remarks and Igor
Soszy\'nski (the referee) for his constructive comments. 
PK gratefully  acknowledges support from Nicolaus Copernicus Center during her stays in 2010 and 2011.
PP is supported from funding to the OGLE project from the European
Research Council under the European Community$'$s Seventh Framework
Programme (FP7/2007-2013)/ERC grant agreement no. 246678, and
by the grant No. IP2010 031570 financed by the Polish Ministry of
Sciences and Higher Education under Iuventus Plus programme.
Part of the research presented in this was supported by the Polish MNiSW grant number N N203 379636.}

\end{document}